# Chaotic Arithmetic Coding for Secure Video Multicast

*Abstract*—Arithmetic Coding (AC) is widely used for the entropy coding of text and video data. It involves recursive partitioning of the range [0,1) in accordance with the relative probabilities of occurrence of the input symbols. A data (image or video) encryption scheme based on arithmetic coding called as Chaotic Arithmetic Coding (CAC) has been presented in previous works. In CAC, a large number of chaotic maps can be used to perform coding, each achieving Shannon optimal compression performance. The exact choice of map is governed by a key. CAC has the effect of scrambling the intervals without making any changes to the width of interval in which the codeword must lie, thereby allowing encryption without sacrificing any coding efficiency. In this paper, we use a redundancy in CAC procedure for secure multicast of videos where multiple users are distributed with different keys to decode same encrypted file. By encrypting once, we can generate multiple keys, either of which can be used to decrypt the encoded file. This is very suitable for video distribution over Internet where a single video can be distributed to multiple clients in a privacy preserving manner.

I. INTRODUCTION (*Heading 1*)

Due to the emergence of the cloud computing paradigm, and he ubiquitousness of portable multimedia-capable devices, the issue of real-time multimedia delivery has gained a lot of importance. The technical challenges involved in such scenarios include providing a compression mechanism that is highly scalable, secure, easily search-able and index-able. All of this should be done that the compression properties are not lost. Providing security in a video communication context is especially challenging, as the security requirements tend to be application and platform-specific, and the input data is characterized by large storage requirements, real-time processing latencies, and the use of standardized video codecs.

Arithmetic coding is extremely efficient for compression efficiency in large data- sets and it achieves the Shannon compression efficiency for large chunks of data [2]. However, conventional implementations are not particularly secure. Thus, arithmetic coding has been interpreted in terms of chaotic maps [11] to be useful for joint compression and encryption of video data[1,12-16].

Many other techniques based on varying the statistical model of entropy coders have been proposed in the research literature [3-6], however these techniques suffer from losses in compression efficiency that result from changes in entropy model statistics and are weak against known attacks [7-10].

Encryption schemes like AES and DES can be used for video data also, but they require specialized hardware for encryption[17-19], whereas these schemes can achieve encryption while compression.

In this paper, we explain how CAC can be used for secure multicast. In typical secure multicast scenario, the content is encoded using different keys and then distributed to multiple users, or the content is not encrypted. With CAC, we can generate multiple keys to decode the same content. Hence, only one encode is required for the entire multicast group and different users can obtain different keys to decode their content.

II. ARITHMETIC CODING WITH PIECE-WISE LINEAR CHAOTIC MAPS

Before we go ahead, we review the basics of chaotic arithmetic coding. We consider a scenario where we have a string $S = x_1, x_2, ...x_N$ consisting of N symbols to be encoded. The probability of occurrence of a symbol $s_i$, $i \in 1, 2, ...n$ is given by $p_i$ such that $p_i = N_i/N$ and $N_i$ is the number of times the symbol $s_i$ appears in the given string S. We next consider a piece-wise linear map ($\rho$) with the following properties:

1. It is defined on the interval [0, 1) to [0, 1) i.e.

    P: [0, 1) → [0, 1)

2. •The map can be decomposed into N piece-wise linear parts $q_k$ i.e.

$$P = \bigcup_{k=1}^{N} q_k$$

3. Each part $q_k$ maps the region on the x axis $[beg_k, end_k)$ to the interval [0, 1) i.e.

    $q_k : [beg_k, end_k) \longrightarrow [0, 1]$

4. The last two propositions lead to:

$$\bigcup_{k=1}^{N} [beg_k, end_k] = [0,1)$$

5. The map $q_k$ is one-one and onto i.e.:

    $\forall X \in [beg_x, end_k)$

    $\exists y \in [0,1): y = q_k(x)$, and

    $\forall y \in [0,1)$

    $\exists x \in [beg_k, end_k): q_k(x) = y$

6. ρ is a many-one mapping from [0, 1) to [0, 1). This implies that the decomposed linear maps (k) don't intersect each other i.e.

$$\forall(k \neq j): [beg_k, end_k] \cap [beg_j, end_j] = 0$$

7. Each linear map qk is associated uniquely with one symbol si. The mapping qk → si is define arbitrarily but the one-one relationship must hold.

8. The valid-input width of each map (qk), given by (endk − begk) is proportional to a probability of occurrence of symbol si.

endk − begk ∝ pi

⇒ endk − begk = C × pi

We recall that $\sum_{k=1}^{N}(end_k - beg_k)$ is same as the input width of $\bigcup_{k=1}^{N}(q_k = \rho)$, which is 1. Also $\sum_{i=1}^{N} p_i = 1$. Thus, we get the value of the constant C to be 1.

$$\Rightarrow end_k - beg_k = p_i$$

Figure 1 shows a sample map fulfilling these properties. Figure 1(a) shows the full map with different parts 1, 2, ...N present, while Figure 1(b) zooms into individual linear part k. The maps are placed adjacent to each other so that each input point is mapped into an output point in the range [0, 1). The total number of distinct ways of arranging N maps to obtain ρ fulfilling the properties mentioned above is given by N! = N × (N − 1) × (N − 2)...3 × 2 × 1, where ! denotes the factorial sign. It is same as arranging these N maps in a sequence, one after another, with the end interval of one map touching the beginning interval of another.

However, there are N different piece-wise maps, each with two possible orienta- tions (with positive or negative slope). Thus, the number of total permutations possible is given by N!2N which is independent of unique symbol probability. Thus, for N-ary arithmetic coding or arithmetic coding with N symbols, it is possible to have N!2N different mappings each leading to same compression efficiency. Since we can arbi- trarily choose any 1 of the N!2N maps, the key space for encoding a single bit of data. It can have a positive or negative slope depending on choice is $[\log_2(N! \, 2^n)]$ bits, where ⌈ ⌉ represents the greatest integer function. For N = 2, it gives 8 mappings. If we increase N to 4 this value increases to 384.

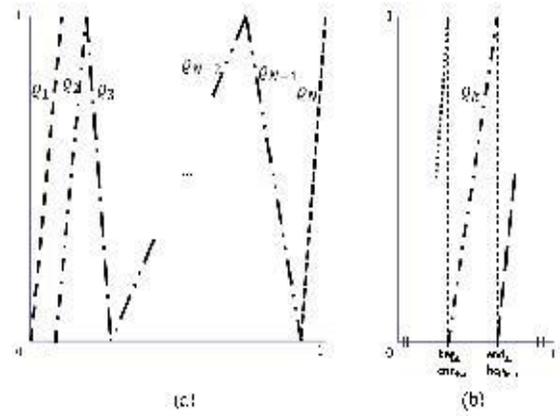

Figure 1: A sample piece-wise linear map for arithmetic coding like compression (a) The entire map is shown (ρ) (b) A single linear part of the map (k) is zoomed

The equation for individual maps can be derived as follows:

$$y' = q_k(x') = \left(\frac{x' - beg_k}{end_k - beg_k}\right) \text{ or } \left(1 - \frac{x' - beg_k}{end_k - beg_k}\right)$$

The equation for the full map is given by

$$y = \rho(x) = q_k(x): beg_{q_k} \leq x < end_k$$

The coding procedure, correspondence to arithmetic coding and compression efficiency of basic chaotic coding is explained in [13]. The compression efficiency of this scheme is explained in [11] to be same as arithmetic coding i.e. it achieves Shannon efficiency for large numbers.

### III. BINARY CHAOTIC ARITHMETIC CODING (BCAC)

In the previous section we explained how arithmetic coding can be viewed as re- iteration on chaotic maps. AC is more commonly implemented in binary mode to reduce the computational requirements of video coders. For same considerations, we discuss implementation and security issues with BCAC in this paper after introducing CAC in last section. The Binary CAC (or BCAC) uses either of the eight equivalent skewed binary maps (shown in Figure 2) based on an input key. These maps differ from each other in the way input is mapped into the chaotic orbit - differ in the interval in which the arithmetic code must lie for a symbol '0' or '1' but the width of inter- val remains the same. In next section, we will formulate a mathematical procedure to generate the eight maps and choose between them using the parameter i.

In the previous section we explained how arithmetic coding can be viewed as reiteration on chaotic maps. AC is more commonly implemented in binary mode to reduce the computational requirement of video coders. For same consideration we discuss implemented and security issues with BCAC in this paper after introducing CAC in last section.

TABLE I. PARAMETER LIST FOR THE EIGHT POSSIBLE CHOICES OF CHAOTIC ENCODER

| Parameter | (a) | (b) | (c) | (d) | (e) | (f) | (g) | (h) |
|---|---|---|---|---|---|---|---|---|
| M1 | p | p | −p | −p | p | −p | −p | p |
| B1 | 0 | 0 | p | p | 1 − p | 1 | 1 | 1 − p |
| M2 | $1-p$ | $p-1$ | $p-1$ | $1-p$ | $1-p$ | $1-p$ | $p-1$ | $p-1$ |
| B2 | $p$ | 1 | 1 | $p$ | 0 | 0 | $1-p$ | $1-p$ |
| N1 | $1/p$ | $1/p$ | $-1/p$ | $-1/p$ | $1/(1-p)$ | $1/(1-p)$ | $-1/(1-p)$ | $-1/(1-p)$ |
| C1 | 0 | 0 | 1 | 1 | 0 | 0 | 1 | 1 |
| N2 | $1/(1-p)$ | $-1/(1-p)$ | $-1/(1-p)$ | $1/(1-p)$ | $1/p$ | $-1/p$ | $-1/p$ | $1/p$ |
| C2 | $-p/(1-p)$ | $1/(1-p)$ | $1/(1-p)$ | $-p/(1-p)$ | $(p-1)/p$ | $1/p$ | $1/p$ | $(p-1)/p$ |
| I1 | 0 | 0 | 0 | 0 | $(1-p)$ | $(1-p)$ | $(1-p)$ | $(1-p)$ |
| I2 | $p$ | $p$ | $p$ | $p$ | 1 | 1 | 1 | 1 |
| I3 | $p$ | $p$ | $p$ | $p$ | 0 | 0 | 0 | 0 |
| I4 | 1 | 1 | 1 | 1 | $1-p$ | $1-p$ | $1-p$ | $1-p$ |
| K | $p$ | $p$ | $p$ | $p$ | $1-p$ | $1-p$ | $1-p$ | $1-p$ |

The Binary CAC (or BCAC) uses either of the eight equivalent skewed binary maps (shown in fig 2) based on a input key. These maps differ from each other in the way of input is mapped into the chaotic orbit. Differ in the interval in which the arithmetic code must lies for the symbol '0' or '1' but the width of interval remain same. In the next section we will formulate a mathematical procedure to generate the eight maps and choose between them using the parameter I.

We define the generalized skewed binary map with the following equations:

$$y = \begin{cases} n_1 x + c_1 & \text{when } x \leq k \\ n_2 x + c_2 & \text{when } x > k \end{cases}$$

Decode $\begin{cases} m_1 y + b_1 & \text{when '0'} \\ m_2 y + b_2 & \text{when '1'} \end{cases}$

Then, the iteration on skewed binary map is defined by the following equation:

$$x = \begin{cases} m_1 y + b_1 & \text{when '0'} \\ m_2 y + b_2 & \text{when '1'} \end{cases}$$

Where n1,n2,c1,c2,m1,m2, b1 and b2 values can be precomputed for different maps and stored in the table for look up for fast acess. Table 1 gives the value of these parameter for all eight chaotic maps.

Grangetto et al. [4] present a Randomized Binary Arithmetic Coding (RBAC) scheme where they change the ordering of '0' and '1' intervals in a Binary Arithmetic Coder (BAC) based on a key. RBAC can be seen as a special case of BCAC where only two of the eight modes of BCAC are used for encryption purposes. Similarly, KSAC [6] can be represented in terms of piece-wiselinear maps by removing the condition of continuity of individual maps (j(x)). Each part jmaps a discontinuous interval on x-axis to the interval [0,1).

IV. SECURE VIDEO MULTICAST

The BCAC schemes allows a given input string to be decoded by more than one key for correct reconstruction. This is attributed to the fact, that for a given bit - two of the eight maps will give the same output interval. For example - encoding '0' with either of the maps given in Figure2(a) and Figure 2(b) will lead to same result. The same is true if we encode '1' with maps given in Figure 2(a) and Figure 2(d).

Thus, we can infer that "Of the 8 possible maps, two maps will give the same reconstruction for a single bit but the choice of bit depends on its value being '0' or '1'." For a N bit message, we have a pool of 2N choices out of the total keyspace of
$2^{3N}$. We illustrate this with the example of a three bit message '001' which is encoded with the maps '1','3' and '6' or simply '136'. For bit '0;, map '2' will decode similar to map '1'. For bit '0', map '4' will decode similar to map '3', and for bit '1', map '5' will decode similar to map '6'. Thus, different users can download the same encoded message and decrypt them on their machine using different keys chosen from the pool of eight keys - '245','246', '235', '236', '135', '136', '145' and '146'.

Thus, the same compressed (and encrypted) data can be sent to all the users in the multicast, while a unique key can be sent to each user. In this example (of three bits compression), the key space is $8^3 = 512$, of which only 8 keys are correct.

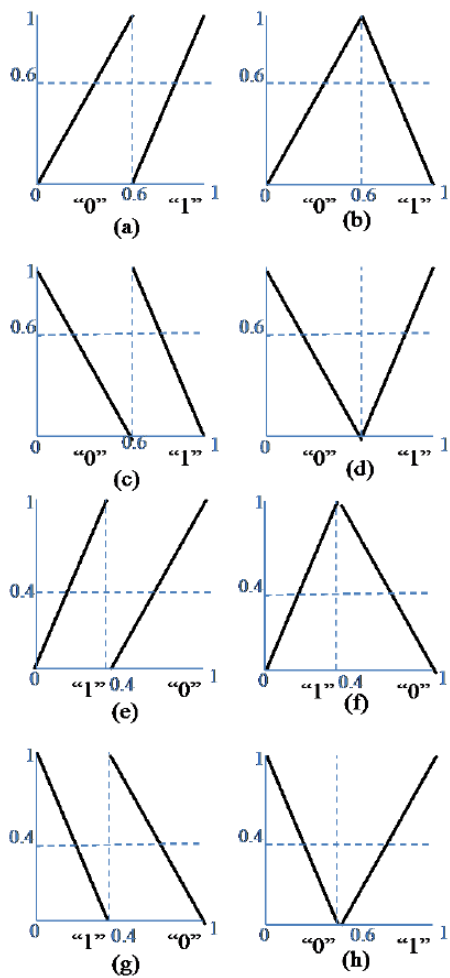

Figure2: Eight modes of Binary Chaotic Arithmetic Coding

Let us say that we choose to encrypt only first M bits of a bitstream of length N. Typically, we may expect N=1000 ∼ 100, 000. However, we can choose small values of M, say M=128. For this choice, we have $2^{128}$ correct decryption keys in a keyspace of $2^{384}$ bits. Thus, the probability of guessing a right key is $2^{-256}$ and a exhaustive search of keyspace will not be helpful. The same compressed video string can be distributed to different users (upto $2^{128}$), each obtaining his own unique to decode/ decrypt it. Identifying the keys can help us to trace users who may be involved in illegal or unauthorized distribution of keys. Therefore, a single encoded stream can be multicasted to different users who can decode video content using their own keys.

If a user relays his key to un-authorized person, as in the case of content piracy, and the un-authorized person is caught for Digital Rights Infringement, the relaying user can be directly identified and blacklisted/ penalized for his improper action. It is not possible to generate a random correct key, as the key is $2^{128}$. But, an attacker may use the knowledge that every three bits of the key are corelated to each other to reduce the attack space to 128. Still, the keyspace is large enough to guarantee success.

*A. Limitations*

A possible limitation of this feature is that we need to send a unique key for each plaintext. In many encryption mechanisms, the same key is used for multiple encodes or session. This is not possible in our scenario, because multiple CAC multicast keys are generated from one original key based on the encoded content. With changing content (plain text), the same original key will lead to different multicast keys. This means that, for every string of N bits (where we encrypt first M bits), we need to transmit 3M bits of key information, to every user in the multicast.

However, by choosing M << N, we can easily overcome this limitation. The overhead of key distribution will be very low. Plus, re-distributing keys adds to the security of multicast system.

Another possible limitation is the possibility of collusion attack. Multiple (say K) users may collude together to mix their keys and generate new keys which are then relayed to illegitimate users. The generated new keys can be different than the original keys. As K increases, it becomes more and more difficult to recognize the original colluding users.

*B. Modification for Secure Multicast*

To address the collusion attack scenario, we propose a modification to the key distri- bution scheme where keys are first encrypted by the coder using an asymmetric en- cryption scheme and then distributed to different users. Thus, we need to provision cryptographic encryption of the keys used by the system. These can be decoded and then used.

In this scenario, we incur the extra cost of cryptographic encryption of the keys (3M bits). By choosing M << N, we can ensure that this overhead is negligible compared to the overall cost of re-encryption for every client.

CONCLUSION

In this paper, we discussed how chaotic arithmetic coding can be used for secure multicast of multimedia data. We found that it is possible for the users to collude if they exchange their keys with each other, making it difficult for DRM authorities to detect the user who leaked the key to unlicensed user. However, we proposed to encrypt these user keys and decode them inside the decoder using a asymmetric system.

REFERENCES

[1] A. Pande, P. Mohapatra, and J. Zambreno, "Using chaotic maps for encrypting image and video content," in Multimedia (ISM), 2011 IEEE International Sympo- sium on. IEEE, 2011, pp. 171–178.

[2] G. Langdon and J. Rissanen, "Compression of black-white images with arithmetic coding," IEEE Trans. Communications, vol. 29, no. 6, pp. 858–867, Jun 1981.

[3] Y. Mao and M. Wu, "A joint signal processing and cryptographic approach to multimedia encryption," IEEE Trans. Image Processing, vol. 15, no. 7, pp. 2061–2075, July 2006.

[4] M. Grangetto, E. Magli, and G. Olmo, "Multimedia selective encryption by means of randomized arithmetic coding," IEEE Trans. Multimedia, vol. 8, no. 5, pp. 905–917, Oct. 2006.

[5] H. Kim, J. Wen, and J. Villasenor, "Secure arithmetic coding," IEEE Trans. Signal Processing, vol. 55, no. 5, pp. 2263–2272, May 2007.

[6] J. Wen, H. Kim, and J. Villasenor, "Binary arithmetic coding withkey-based interval splitting," IEEE Trans. Signal Processing Letters, vol. 13, no. 2, pp. 69–72, Feb. 2006.

[7] G. Jakimoski and K. Subbalakshmi, "Cryptanalysis of some multimedia encryption schemes," IEEE Trans. Multimedia, vol. 10, no. 3, pp. 330–338, April 2008.

[8] J. Zhou, O. C. Au, X. Fan, and P. H. W. Wong, "Joint security and performance enhancement for secure arithmetic coding," in ICIP, 2008, pp. 3120–3123.

[9] J. Zhou, O. C. Au, P. H. Wong, and X. Fan, "Cryptanalysis of secure arithmetic coding," in ICASSP, 2008, pp. 1769–1772.

[10] H.-M. Sun, K.-H. Wang, and W.-C. Ting, "On the security of the secure arithmetic code," IEEE Trans. Information Forensics and Security, vol. 4, no. 4, pp. 781–789, 2009.

[11] N. Nagaraj, P. Vaidya, and K. Bhat, "Arithmetic coding as a non-lineardynam- ical system," Communications in Nonlinear Science and Numerical Simulation, vol. 14, no. 4, pp. 1013–1020, 2009.

[12] A Pande, J Zambreno, P Mohapatra "Joint Video Compression and Encryption UsingArithmetic Coding and Chaos", IEEE International Conference on Internet Multimedia *Systems Architecture and Applications*, December 17-19, 2010, Bangalore, India.

[13] A. Pande, J. Zambreno, and P. Mohapatra, "Joint video compression and encryp- tion using arithmetic coding and chaos," in IEEE International Conference on Internet Multimedia Systems Architecture and Application, 2010.

[14] A. Pande, P. Mohapatra, J. Zambreno, "Securing Multimedia Content using Joint Compression and Encryption", IEEE Multimedia, Vol. 20, no. 4, Oct-Dec 2013.

[15] A Pande, J Zambreno, P Mohapatra "Hardware Architecture for Simultaneous Arithmetic Coding and Encryption" The International Conference on Engineering of Reconfigurable Systems and Algorithms (ERSA), July 2011, Las Vegas, USA.

[16] A Pande, J Zambreno, P Mohapatra "Architectures for Simultaneous Coding and Encryption Using Chaotic Maps" *IEEE International Symposium on VLSI (ISVLSI)*, pp. 351-352, July 2011, Chennai, India.

[17] J. Zambreno, D. Nguyen, and A. N. Choudhary, "Exploring area/delay tradeoffs in an AES FPGA implementation," in Proc. IEEE Intl. Conf. Field Programmable Logic and Applications, FPL 2004, 2004, pp. 575–585.

[18] FIPS 197, "Announcing the Advanced Encryption Standard," Nov 2001.

[19] FIPS 46-2, "Announcing the standard for Data Encryption Standard," Dec 1993.